\documentclass[11pt,a4paper,leqno]{article}
\usepackage{graphicx}
\usepackage{a4wide}
\usepackage{amsmath}
\usepackage{color}
\usepackage{amssymb}
\usepackage{amsbsy}
\usepackage{bm}
\graphicspath{{Figures/}}

\begin{document}

\title{Time-evolution of the ion velocity distribution function in the discharge of a Hall effect thruster}

\author{S. Mazouffre$^\dagger$, D. Gawron$^\dagger$, N. Sadeghi$^\ddagger$,}

\maketitle
$^\dagger$ {\small ICARE, CNRS, 1C avenue de la Recherche Scientifique, 45071 Orl\'{e}ans, France.}\\
$^\ddagger$ {\small LSP, Joseph Fourier University - CNRS, 140 Av.
de la Physique, 38402 St Martin d'H\`eres, France.}

\begin{abstract}
The temporal characteristics of the Xe$^+$ ion axial Velocity
Distribution Function (VDF) were recorded in the course of
low-frequency discharge current oscillations ($\sim$~14\,kHz) of the
5\,kW-class PPS$\circledR$X000 Hall thruster. The evolution in time
of the ion axial velocity component is monitored by means of a laser
induced fluorescence diagnostic tool with a time resolution of
100\,ns. As the number of fluorescence photons is very low during
such a short time period, a hom-made pulse-counting lock-in system
was used to perform real-time discrimination between background
photons and fluorescence photons. The evolution in time of the ion
VDF was observed at three locations along the thruster channel axis
after a fast shut down of the thruster power. The anode discharge
current is switched off at 2\,kHz during 5\,$\mu$s without any
synchronization with the current oscillation cycle. This approach
allows to examine the temporal behavior of the ion VDF during decay
and ignition of the discharge as well as during forced and natural
plasma oscillations. Measurements show that the distribution
function of the axial component of the Xe$^+$ ion does change
periodically in time with a frequency close to the current
oscillation frequency in both forced and natural cases. The ion
density and the mean velocity are found to oscillate whereas the
velocity dispersion stays constant, which indicates that ionization
and acceleration layers have identical dynamics. Finally, variations
over time of the electric field are for the first time
experimentally evidenced in a crossed-field discharge.
\end{abstract}

\mbox{}\newline
\newline
\noindent Submitted to {\it Physics of Plasmas}

\newpage
\section{Introduction}

A Hall Effect Thruster (HET) is a gridless ion accelerator that
finds applications in the field of spacecraft
propulsion~\cite{JPL,MARTINEZ}. Such a type of electric propulsive
device is especially suited for long duration missions and for
maneuvers that require a large velocity increment. HETs are at
present mostly employed for geostationary communication satellite
orbit correction and station keeping. Other fields of application
are envisaged for the near future. Low power Hall thrusters seem
suited for drag compensation of observation satellites that operate
on a low-altitude Earth orbit. The use of high power Hall thrusters
for orbit raising and orbit topping maneuvers of communication
satellites would offer significant benefits in terms of launch mass,
payload mass and operational life. In addition, large Hall thrusters
appear as good candidates to be used as the primary propulsion
engine for robotic space probes during interplanetary journeys
towards far-off planets and asteroids.

\begin{figure}[t]
\centering
\includegraphics[width=8cm]{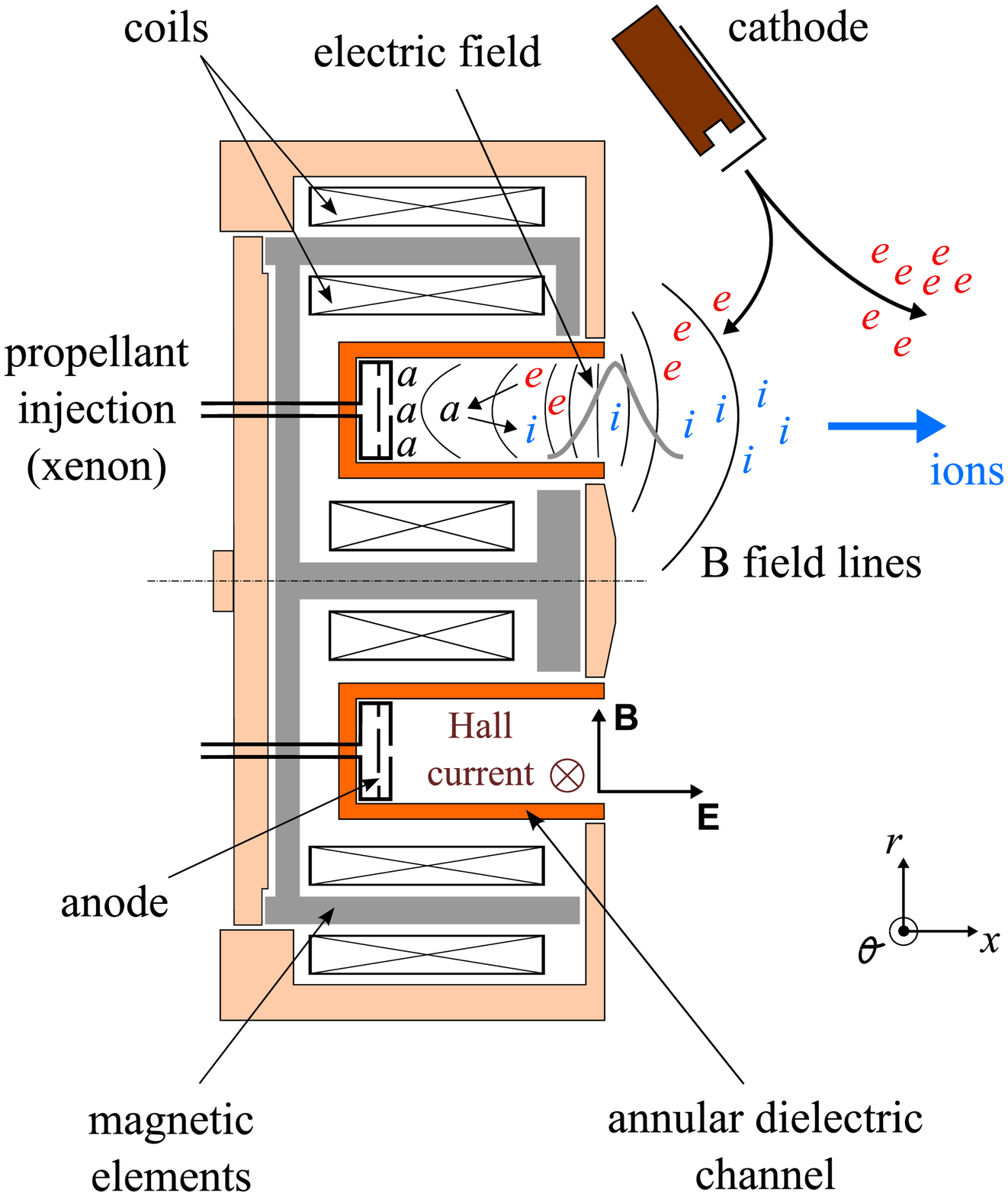}
\caption{\small{Cross-section view of a Hall effect thruster. The
symbol {\it e} stands for electron, {\it a} for atom and {\it i}
 for ion. The channel exit plane is referred to as $x$~=~0 in this work.}} \label{fig:HET}
\end{figure}

A Hall effect thruster is a low-pressure DC discharge in crossed
electric and magnetic fields configuration~\cite{MARTINEZ,
ZHURIN,KIM}. Xenon is generally used as a propellant gas due to its
high atomic mass and low ionization energy. A schematic of a HET is
depicted in Fig.~\ref{fig:HET}. The anode is located at the upstream
end of a coaxial annular dielectric channel that confines the
plasma. The cathode is situated outside. A set of coils combined
with magnetic parts provide a radially directed magnetic field {\bf
B} of which the strength is maximum in the vicinity of the channel
exhaust. The magnetic field is chosen strong enough to make the
electron Larmor radius much smaller than the channel characteristic
sizes, but weak enough not to affect ion trajectories. The gas
injected through the anode is ionized inside the channel by electron
impacts. As the magnetic field considerably slows down the electron
motion towards the anode, the applied potential concentrates in a
restricted area at the channel entrance. The corresponding axial
electric field {\bf E} then accelerates ions out of the channel,
which generates thrust. The ion beam is neutralized by a fraction of
electrons emitted from the cathode. The crossed {\bf E} and {\bf B}
geometry is at the origin of a large electron azimuthal drift --~the
Hall current~-- that is responsible for the efficient ionization of
the supplied gas. When operating near 1.5~kW, a HET ejects ions at
20\,km\,s$^{-1}$ and generates 100~mN of thrust with an overall
efficiency of about 50~\%~\cite{GASCON}.

It is well-established that the crossed-fields discharge of a Hall
effect thruster is strongly non-stationary~\cite{ZHURIN,CHOUEIRI}.
This specific type of magnetized plasma displays numerous types of
oscillations, which encompass many kinds of physical phenomena, each
with its own length and time scales~\cite{CHOUEIRI}. Current and
plasma fluctuations, of which the frequency range stretches from
$\sim$\,10\,kHz up to $\sim$\,100\,MHz, play a major role in
ionization, particle diffusion and acceleration processes.
Low-frequency plasma oscillations in the range 10-30\,kHz, so-called
breathing oscillations, are especially of interest as they carry a
large part of the power~\cite{JACEK}. Breathing oscillations find
their origin in a prey-predator type mechanism between atoms and
ions as shown by Boeuf and co-workers~\cite{BOEUF}. In short, these
oscillations originate in a periodic depletion and replenishment of
the neutrals near the exhaust of the thruster channel due to the
efficient ionization of the gas. The frequency is then linked to the
time it takes for atoms to fill in the ionization region. With an
atom thermal speed of 300\,m/s and a region size of 20\,mm, one
finds a frequency of 15\,kHz. The breathing phenomenon not only
disturbs the discharge current but also has a strong impact on
several quantities. The plume shape and the ion beam divergence
change during an oscillation cycle of the current as was shown by
means of CCD imaging~\cite{VANESSA}. A time variation of the ion
beam energy in a Hall thruster far field was evidenced using a
retarding potential analyzer~\cite{ANDRE}. The electron density and
the plasma potential oscillate at low frequency~\cite{LUC}.
Conversely, the electron temperature stays unchanged. Oscillations
of the aforementioned quantities are most likely connected with a
time variation of the potential distribution or, in other words,
with the variation with time of the accelerating electric field.
Therefore, it appears of great interest to investigate the temporal
behavior of the electric field that certainly hides a rich and
intricate dynamics. Across the acceleration layer, the medium is
collisionless, i.e. ion-ion and ion-atom collision events are
scarce. As a consequence, the electric field can be directly
inferred from the Xe$^+$ velocity which then becomes the quantity to
be examined.

Laser Induced Fluorescence (LIF) spectroscopy in the near infrared
has often been used in the past few years to measure the
time-averaged velocity distribution function (VDF) of Xe$^+$ ions in
the plasma of a Hall effect
thruster~\cite{HARGUS,HARGUS2,GAWRON,MAZ1}. LIF spectroscopy is a
non intrusive diagnostic tool that enables an accurate determination
of the local velocity of atoms along the laser beam direction by
measuring the Doppler shift of absorbed photons. The metastable
Xe$^+$ ion VDF is recorded by collecting fluorescence radiation at
541.9\,nm after excitation of the 5d\,$^2$F$_{7/2} \rightarrow$
6p\,$^2$D$^o_{5/2}$ transition at $\lambda$ =
834.7233\,nm~\cite{GAWRON}. A phase sensitive detection method is
often used to capture the fluorescence signal. However, this method,
which is powerful enough for the extraction of a signal in an
environment with a high background noise level, offers a poor time
resolution. To achieve the measurement of the time-resolved Xe$^+$
ion VDF in the plasma of a Hall effect thruster, it is necessary to
develop a bench able to detect LIF photons with a time resolution
around 1$\mu$\,s. In normal operating conditions for a HET, the
number of fluorescence photons observed at 541.9 nm with a
continuous laser beam tuned at 834.7233\,nm with about 1\,mW/mm$^2$
power density is on the order of 10$^{-2}$ per $\mu$s. Under
identical experimental conditions, the number of background photons
generated by the plasma at 541.9 nm during 1\,$\mu$s is typically 1,
which means a ratio of 100 between the two signal amplitudes. The
laser system must therefore be able (i) to detect a tiny amount of
photons hidden in a strong background (ii) to determine with a high
accuracy the exact moment in time fluorescence photons have been
produced. One must therefore turn to a photon-counting technique.

In this contribution, we present time-resolved measurements of the
Xe$^+$ ion axial VDF in the discharge of the 5\,kW-class
PPS$\circledR$X000 Hall thruster fired at 500\,V discharge voltage
and 6\,mg/s xenon mass flow rate. The evolution in time of the VDF
was recorded at several locations during the transient regime that
follows a fast anode discharge current ignition in order to
investigate the ion dynamics during forced and free low frequency
current oscillations. The current is switched off during 5\,$\mu$s
at a 2\,kHz repetition rate without any synchronization with the
discharge current oscillation cycle. The outline of this paper is as
follows. In Sec.~\ref{SEC_DIAG}, the LIF optical assembly is
described and the pulse-counting technique is introduced.
Section~\ref{SEC_VDF} shows contour plots of the time-varying Xe$^+$
ion VDF as well as traces of various velocity groups at the thruster
outlet. In Sec.\ref{SEC_MACRO}, the temporal behavior of macroscopic
quantities like the density, the mean velocity and the velocity
dispersion, is examined and discussed. Section~\ref{SEC_EFIELD}
reports on variation over time of the accelerating electric field in
the crossed-field discharge of a Hall thruster. Finally, concluding
remarks will be presented in Sec.~\ref{SEC_CONCLU}.

\section{Diagnostic technique}
\label{SEC_DIAG}

\subsection{Optical bench and collection branch}

The LIF optical assembly is extensively described in
Ref.~\cite{MAZ1}. The laser beam used to excite Xe$^+$ metastable
ions at 834.7\,nm is produced by an amplified tunable single-mode
external cavity laser diode. The wavelength is accurately measured
by means of a calibrated wavemeter whose absolute accuracy is better
than 100 MHz, which corresponds to 90\,m/s. A plane scanning
Fabry-Pérot interferometer with a 1.29\,GHz free spectral range is
used to real-time check the quality of the laser mode and to detect
mode hops. The primary laser beam is modulated by a mechanical
chopper at a low frequency $\sim$\,20\,Hz before being coupled into
a 50\,m long optical fiber of 50\,$\mu$m core diameter. The fiber
output is located behind the thruster. Collimation optics are used
to form a narrow beam that passes through a small hole located at
the back of the PPS{\circledR}X000 thruster. The laser beam
propagates along the channel axis in the direction of the ion flow.
Typically, the laser power density reaches 3\,mW/mm$^2$, which
warrants a weak saturation effect on the studied transition.

A collection branch made of a 40\,mm focal length lens, which
focuses the fluorescence light onto a 200\,$\mu$m core diameter
optical fiber, is mounted onto a travel stage perpendicular to the
channel axis. The magnification ratio is 1, meaning that the spatial
resolution is 200\,$\mu$m in axial direction. A 16\,mm long slit was
made in the channel dielectric outer wall in order to carry out
measurements inside the channel. The fluorescence light transported
by the 200\,$\mu$m fiber is focused onto the entrance slit of a
20\,cm focal length monochromator that isolates the 541.9\,nm line
from the rest of the spectrum. A photomultiplier tube serves as a
light detector.

\subsection{Lock-in photon-counting device}

The pulse (or photon) counting technique allows the detection of a
very low level signal with an excellent time resolution. When
combined with a modulation of the laser light intensity, the pulse
counting technique can distinguish between LIF photons and
spontaneous emission photons. The technique is known as the
time-resolved pulse counting lock-in detection technique. In this
work, a customized pulse-counting system is used to measure
time-dependent ion VDF~\cite{NADER}. Here we briefly outline the
main characteristics and settings of the system.

\begin{figure}[t]
\centering
\includegraphics[width=12cm]{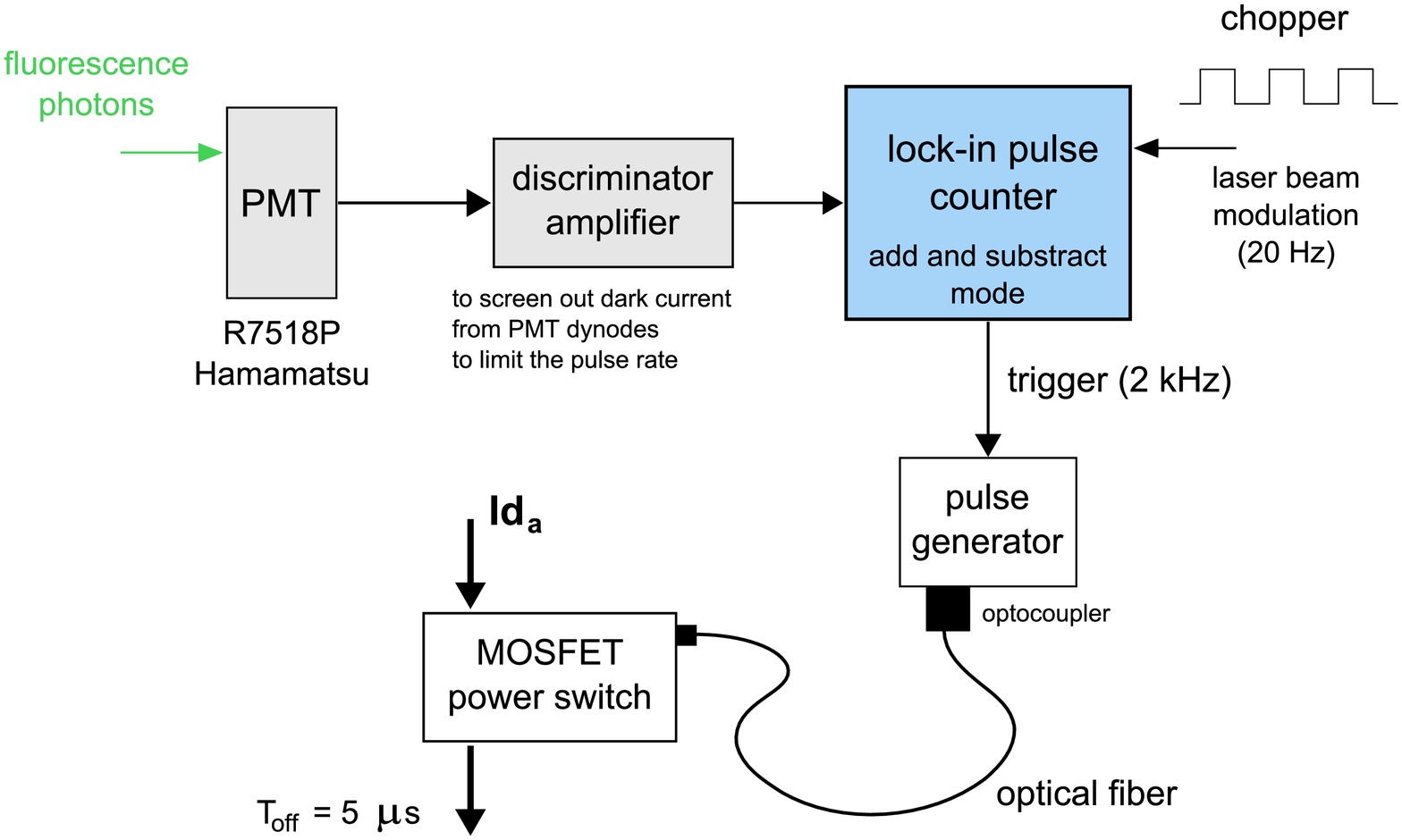}
\caption{\small{Block diagram of the lock-in pulse counting system
used in this work to measure the time-resolved Xe$^+$ ion VDF by
means of LIF spectroscopy. The PMT is placed behind a 20 cm focal
length monochromator. The anode discharge current switch is
externally driven by the counter.}} \label{fig:COUNTER}
\end{figure}

A block diagram of the pulse counting system is shown in
Fig.~\ref{fig:COUNTER}. Photons are detected by means of a high gain
and low dark noise PMT (R7518-P from Hamamatsu). A fast amplifier
and discriminator module (9302 from Ortec - 100 MHz counting rate)
is used to screen out dark current from PMT dynodes, to limit the
pulse rate thereby avoiding saturation of the pulse counter, and to
transform any single event --~here the arrival of a photon~-- into
either a NIM or a TTL pulse. Pulses are subsequently treated by the
lock-in pulse counter device, which counts events as a function of
time. A trigger starts the counter which segments photon count data
into sequential time bins. Notice that up to 32 kbins are available.
The width of the bins can be set from 10\,ns to 655\,s. The
instrument records the number of photons that arrive in each bin. In
order to greatly improve the signal-to-noise ratio, the counter is
able to operate in real-time addition-subtraction mode. The laser
beam intensity is modulated at low frequency ($\sim$\,20\,Hz) by
means of a mechanical chopper. Each pulse recorded when the laser is
propagating through the plasma (laser-on mode) is added to the time
series; the signal corresponds to LIF photons plus background
photons. Each pulse recorded when the laser is suspended (laser-off
mode) is subtracted from the time series: in that case the signal is
solely composed of background photons. A 2\,kHz trigger signal
generated by the counter itself was used to define the start of the
measurement cycle. The time resolution, i.e. the width of each bin,
was set to 100\,ns and 5000 bins were used. The duration of one
measurement cycle is therefore 500\,$\mu$s, corresponding to about 6
times the period of low-frequency current oscillations of the
PPS$\circledR$X000 thruster operating at 500\,V and 6\,mg/s. In
order to obtain a reasonable
signal-to-noise ratio, light was accumulated over 1 million cycles.\\
\newline
\noindent The procedure to obtain the time-resolved ion VDF is the
following:
\begin{itemize}
\item the laser is fixed at a given wavelength $\lambda$ corresponding to
a certain ion velocity group $\delta v$. The extent (dispersion) of
the velocity group results from the spectral width of the laser beam
and from the thermal expansion of the laser cavity. A feedback loop
allows a minimization of the shift of the laser wavelength. The
dispersion is thus around 10\,m/s.
\item The pulse counter is used to record the
number of fluorescence photons induced by excitation of metastable
ions at $\lambda$. That means we follow the temporal evolution of
the velocity group $\delta v$.
\item After about one million discharge current disconnection cycles the
laser wavelength is changed and a new measurement starts.
\end{itemize}
To obtain a smooth ion VDF about 15 to 20 different wavelengths are
used.

\section{Temporal characteristics of the ion VDF}
\label{SEC_VDF}

\subsection{Experimental conditions - Ion emission light}

All measurements were performed in the PIVOINE-2g test-bench. The
PPS$\circledR$X000 Hall effect thruster was equipped with BN-SiO$_2$
channel walls and with a carbon anode. All thruster parameters were
kept unchanged during the experiments: The applied voltage $U_d$ was
set to 500\,V, the anode xenon mass flow rate $\Phi_a$ was fixed at
6\,mg/s and the magnetic field strength $B$ was $\sim$~150\,G. The
mean discharge current is 5.4\,A and the oscillation frequency is
found to be 13.7\,kHz. The temporal characteristics of the ion VDF
were recorded for three locations along the channel centerline. Each
position defines a distinct area in terms of electric field
magnitude~\cite{MAZ1}. The position $x$~=~-15\,mm corresponds to a
zone through which the electric field is almost zero and the
ionization is strong. At $x$~=~-2.5\,mm, the electric field is large
and ionization is maintained. The channel exit plane, $x$~=~0\,mm,
corresponds to a region where ionization ceases and the electric
field strength is high with $E \approx$~300\,V/cm.

In this study, the ion flow dynamics was investigated before and
after a fast shut-down of the anode discharge current. It was
therefore possible to examine the temporal characteristics of the
ion VDF during plasma breakdown and ignition as well as during
forced and free current oscillations. The anode current is switched
off during 5\,$\mu$s at 2\,kHz by way of an optically controlled
fast power switch based on a MOSFET~\cite{VANESSA}. The power switch
is directly driven by the counter, as shown in
Fig.~\ref{fig:COUNTER}. There is no synchronization between the
power cut cycle and the discharge current waveform. In other words,
the anode current is switched off randomly at any time with respect
to the discharge current natural oscillations.

\begin{figure}[t]
\centering
\includegraphics[width=12cm]{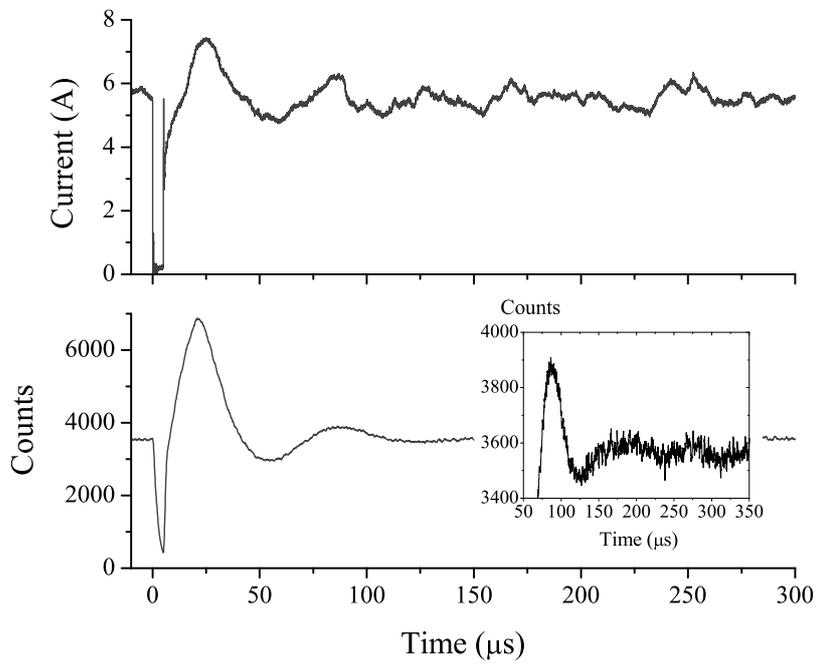}
\caption{\small{(Top) Snapshot of the anode current waveform. The
discharge current is switched for 5\,$\mu$s at $t$~=~0\,$\mu$s.
(Bottom) Change in time of the ion emission at 541.9\,nm observed
using the photon counting technique at $x$~=~-2.5\,mm. The inset
panel displays enlargement of the emission signal from 50\,$\mu$s
until 350\,$\mu$s. The oscillation frequency is $\sim$\,12\,kHz.}}
\label{fig:EMISSION}
\end{figure}

Figure~\ref{fig:EMISSION} displays the intensity of the 541.9\,nm
ion line as a function of time observed at $x$~=~-2.5\,mm recorded
with the photon-counting device operating in addition mode only. A
snapshot of the anode current waveform is also shown. Measurement
reveals the time evolution of natural plasma emission. The power is
always switched off at $t$~=~0\,$\mu$s (reference time). The plasma
decays with a 1/$e$ time of 2.1\,$\mu$s. However, it does not fully
vanish as charged-particle recombination and diffusion processes
timescales are not infinitely shorter compared to the 5\,$\mu$s
current-off time period. On the contrary, both the anode current and
the Hall current cancel in a few hundreds of ns~\cite{IEPC2003}.
This point is specifically addressed in the next sections. A large
amount of light is quickly produced at the re-ignition stage, as can
be seen in Fig.~\ref{fig:EMISSION}. In like manner, a discharge
current burst always occurs at re-ignition~\cite{VANESSA,IEPC2003}.
This forced plasma oscillation originates in the sudden ionization
of the great amount of propellant atoms accumulated inside the
channel when the discharge is off. The plasma oscillates with a mean
period of about 83\,$\mu$s that corresponds to a 12\,kHz frequency.
As can be observed in Fig.~\ref{fig:EMISSION}, the amplitude of
light oscillations diminishes with time, and the signal finally
approaches a constant level. This phenomenon is due to the fact
that, the current disconnection cycle is not synchronized with the
natural current oscillations, while data acquisition is a cumulative
process over thousand of cycles.

\subsection{Contour map}

\begin{figure}[t]
\centering
\includegraphics[width=14cm]{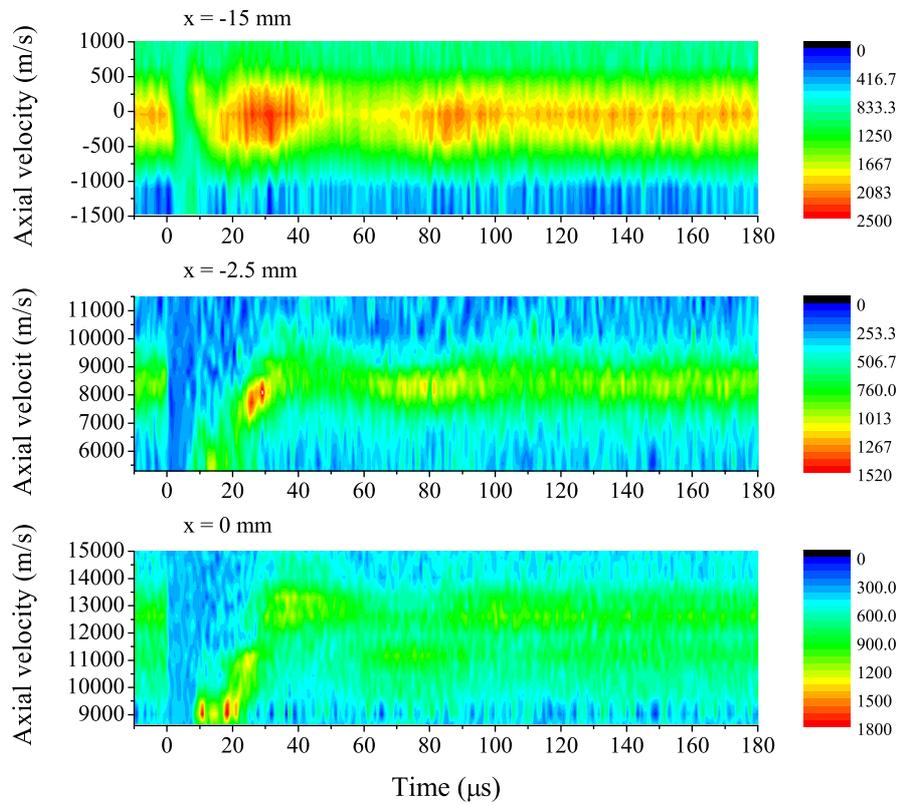}
\caption{\small{Contour map of the Xe$^+$ ion axial VDF as a
function of time for three locations along the channel axis of the
PPS$\circledR$X000 Hall thruster fired at 500\,V: $x$~=~-15\,mm,
$x$~=~-2.5\,mm and $x$~=~0\,mm.}} \label{fig:MATRIX}
\end{figure}

Figure~\ref{fig:MATRIX} shows contour plots of the time evolution of
the Xe$^+$ ion velocity distribution functions for three positions
along the thruster channel axis, respectively $x$~=~-15\,mm,
$x$~=~-2.5\,mm, and $x$~=~0\,mm. All velocity groups vanish quickly
after the current is switched off on a time scale on the order of a
few 100\,ns. This property indicates the electric field cancels
almost instantaneously when the current is stopped. As a
consequence, the discharge current as well as the Hall current
disappear over an extremely short time period, as experimentally
observed~\cite{IEPC2003}. Yet, very slow ions do not fully disappear
in 5\,$\mu$s as recombination, ambipolar diffusion to walls and
drift out of the acceleration zone are slow processes. In
Fig.~\ref{fig:MATRIX}, the fluorescence signal is significantly
above zero after 5\,$\mu$s for $x$~=~-15\,mm. The remark holds also
true for the ion emission signal, see Fig.~\ref{fig:EMISSION}. Note
that a Xe$^+$ ion travels 1.5\,mm in 5\,$\mu$s at the thermal speed
$v_{\rm th} \approx$~300\,m/s. Very slow ions, i.e ions with the
atom speed, are always produced first at re-ignition as
electrostatic acceleration is not an instantaneous process. In
Fig.~\ref{fig:MATRIX}, slowest ions are indeed observed first and
the mean velocity gradually increases up to a limit. Ions moving
with a velocity close to the thermal speed are not visible in
Fig.~\ref{fig:MATRIX} for $x$~=~-2.5\,mm and $x$~=~0\,mm as the ion
VDF is truncated due to a lack of data points. The large production
of ions at plasma ignition originates from the fact that the channel
is entirely filled up with xenon atoms, as previously explained. As
can be seen in Fig.~\ref{fig:MATRIX}, the Xe$^+$ ion VDF in axial
direction changes in time, especially during the first two plasma
oscillations. This phenomenon indicates that the acceleration
potential, and thus the electric field, is likely to vary in time
with a frequency on the order of the main discharge current
oscillation frequency.

\subsection{Velocity groups}

\begin{figure}[t]
\centering
\includegraphics[width=14cm]{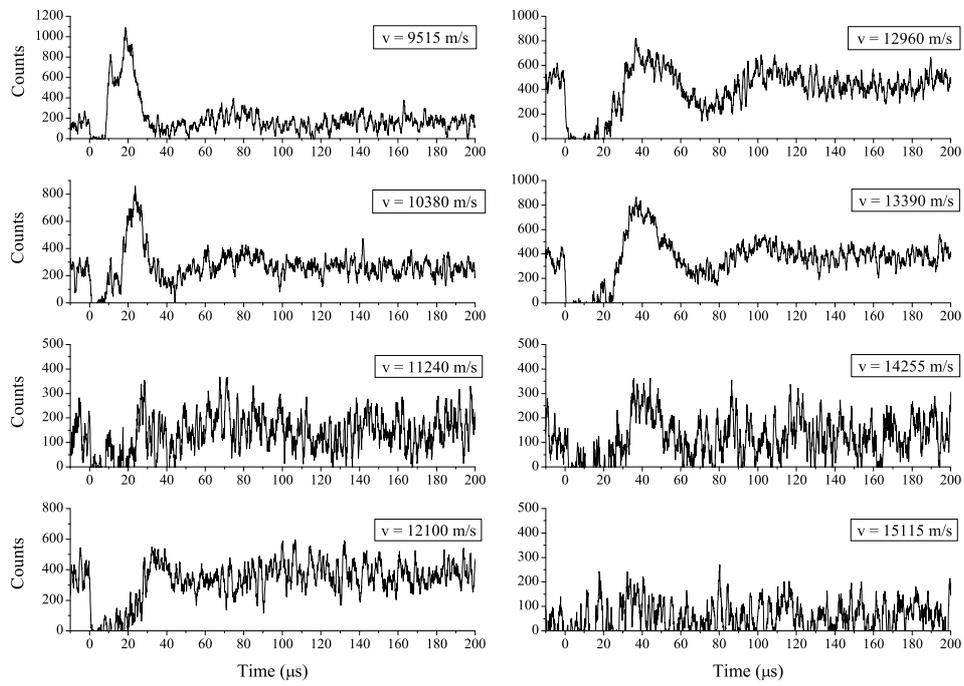}
\caption{\small{Trace of the time evolution of eight ionic velocity
groups at the channel exit plane ($x$ = 0 mm).}} \label{fig:GROUPS}
\end{figure}

Examination of the temporal characteristics of the ion VDFs provides
a general outline of the ion dynamics in the discharge of a Hall
thruster. In contrast, a critical analysis of the temporal behavior
of ion velocity groups reveals in great detail the intricate
character of the physics at work. The time evolution at the channel
outlet of eight well-identified ionic velocity groups $\delta v$ is
shown in Fig.~\ref{fig:GROUPS}. The temporal evolution is monitored
at the thruster channel exhaust. Graphs correspond to horizontal
cross-sections of the lower contour plots in Fig.~\ref{fig:MATRIX}.
All velocity groups quickly vanish as soon as the power is
switched-off: the mean 1/$e$ decay time for all $\delta v$ is
$\sim$\,1.5\,$\mu$s. The discharge current as well as the Hall
current decay in about the same time period~\cite{IEPC2003}.
Nonetheless, fastest ions disappear out of the acceleration region
first: the 1/$e$ decay time is 1.1\,$\mu$s at 12960\,m/s. On the
contrary, slowest ions are produced first when the discharge current
is re-ignited. In Fig.~\ref{fig:GROUPS}, the group $\delta
v$~=~9515\,m/s is detected at 9\,$\mu$s and it reaches its highest
amplitude at 19\,$\mu$s whereas the group $\delta v$~=~13390\,m/s is
detected at 20\,$\mu$s and it goes through a maximum at 39\,$\mu$s.
All velocity families oscillate nonetheless with the same period of
time $T \approx$~73\,$\mu$s, that means with a frequency around
14\,kHz.

When the power is turned on again, ionization immediately takes
place, see Fig.~\ref{fig:EMISSION}. Discharge and Hall current are
restored in less than 1\,$\mu$s and ions are gradually accelerated
as shown in Fig.~\ref{fig:GROUPS}. All these facts indicate that the
electric field is established on a microsecond timescale at
re-ignition. However, it seems at first sight in manifest
contradiction with the time it takes to detect fast ions inside the
thruster channel. The time of flight of a given ion velocity group
across the acceleration layer after plasma re-ignition can be
assessed by numerically solving the particle motion equation in an
external electric field $E$:
\begin{equation}
\label{motion} dv = \frac{e}{m} \, E \, dt,
\end{equation}
where $e$ is the elementary charge and $m$ is the mass of a xenon
ion. For simplicity's sake, the electric field is first set constant
in space and in time through the whole acceleration region (from
$\sim x$~=~-10\,mm to 20\,mm) with a magnitude of
170\,V/cm~\cite{MAZ1}. Moreover, the field is assumed to be
instantaneously created and the ionization process is stationary and
homogeneous. The initial velocity $v_0$ is fixed as the thermal
speed $v_{\rm th}$. Computations reveal that the group $\delta
v$~=~13390\,m/s appears at the exit plane at $t$~=~1.05\,$\mu$s.
Ions have then travelled 7.2\,mm, a distance compatible with the
acceleration layer size $L \sim$\,30\,mm. However, this velocity
group is observed first at $t = 20\,\mu$s, see
Fig.~\ref{fig:GROUPS}. Another approach consists of taking a steady
electric field distribution similar to the measured one~\cite{MAZ1}.
When ions are created about 19\,mm inside the channel, computations
indicate that the group $\delta v$~=~13390\,m/s is indeed observed
at $x$~=~0\,mm for $t \approx 20\,\mu$s. Yet ions moving
at~9515\,m/s are solely seen a few hundreds of ns earlier, in
contradiction with experimental outcomes.

\begin{figure}[t]
\centering
\includegraphics[width=12cm]{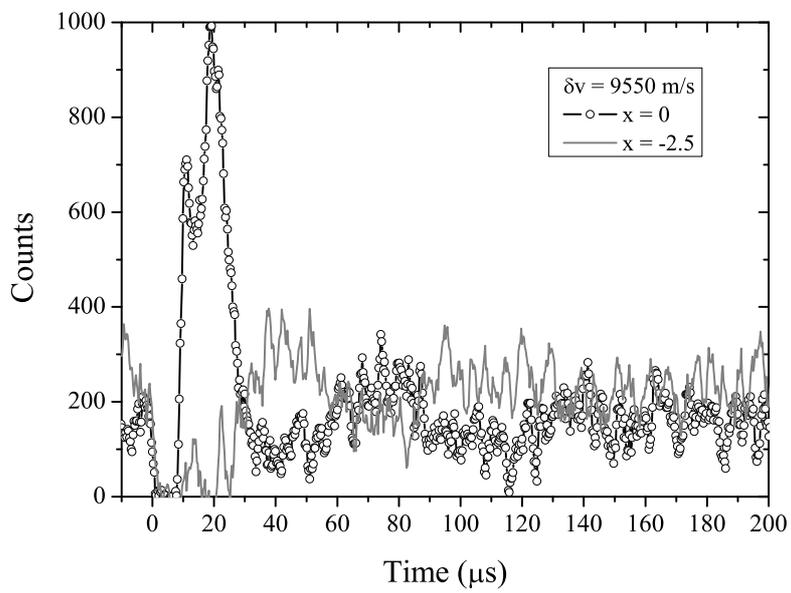}
\caption{\small{Time evolution of the velocity group $\delta v
\approx$~9550\,m/s for two locations: $x$~=-2.5 mm and $\delta
v$~=~9605\,m/s (solid line) and $x$~=~0 mm and $\delta v$=~9515\,m/s
(circle).}} \label{fig:COMPARVDF}
\end{figure}

One way to better duplicate reality consists for instance of
considering that the ionization profile or the electric field can
evolve in the course of time.
With a stationary electric field profile similar to the measured
one~\cite{MAZ1}, numerical simulations using Eq.~\ref{motion} show
that ions must be created first in the vicinity of the channel
outlet and the ionization front must move towards the anode with a
speed on the order of 500\,m/s to reproduce trends that are
experimentally observed in Fig.~\ref{fig:GROUPS}. Propagation of an
ionization wave through the acceleration layer was proposed to
explain experimental results acquired by means of time-resolved
optical emission spectroscopy with a fiber comb~\cite{ANDRE} as well
as CCD images of the plume behavior with
a speed around 1-2\,km/s.\\
\newline
\noindent Figure~\ref{fig:COMPARVDF} shows the time evolution of the
velocity group $\delta v$~=~9550\,m/s for two locations,
respectively $x$~=~-2.5\,mm and $x$~=~0\,mm. The ion family is first
observed at the channel exit plane at $t = 8\,\mu$s before being
detected inside the channel at $t \approx 25\,\mu$s. The highest
amplitude is attained for $t = 19\,\mu$s and $t = 45\,\mu$s at
$x$~=~0\,mm and $x$~=~-2.5\,mm, respectively. Results indicate that
ions are created within a broad region inside the channel: ions
created close to the exit plane are indeed detected first.
Nevertheless, experimental outcomes cannot be correctly simulated
when assuming that an ionization wave travels from the plume near
field towards the anode while the electric field distribution stays
unchanged. On the contrary, results suggest that the ionization
front as well as the acceleration layer change in time together.
Numerical simulations, though basic, and experimental data are
therefore in favor of complex plasma dynamics within the discharge
of a Hall thruster. It is worth noticing that the behavior of Xe$^+$
ion velocity groups during a forced oscillation that follows a
power-off period may not directly image the normal behavior during a
natural plasma oscillation from the viewpoint of amplitude of the
observed phenomena.

\section{Evolution in time of the density, mean velocity and dispersion}
\label{SEC_MACRO}

\begin{figure}[t]
\centering
\includegraphics[width=12cm]{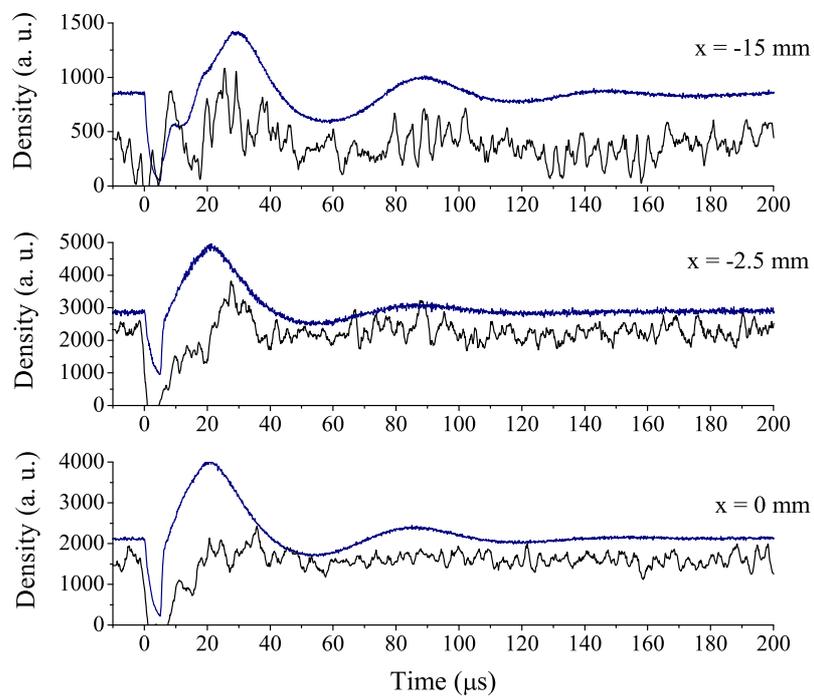}
\caption{\small{Time evolution of the metastable Xe$^+$ ion relative
density for three positions along the channel axis of the
PPS$\circledR$X000 Hall thruster. The density is given by the VDF
area. Also shown is the emission profile at 541.9 nm (blue line).}}
\label{fig:DENSITY}
\end{figure}

\begin{figure}[t]
\centering
\includegraphics[width=12cm]{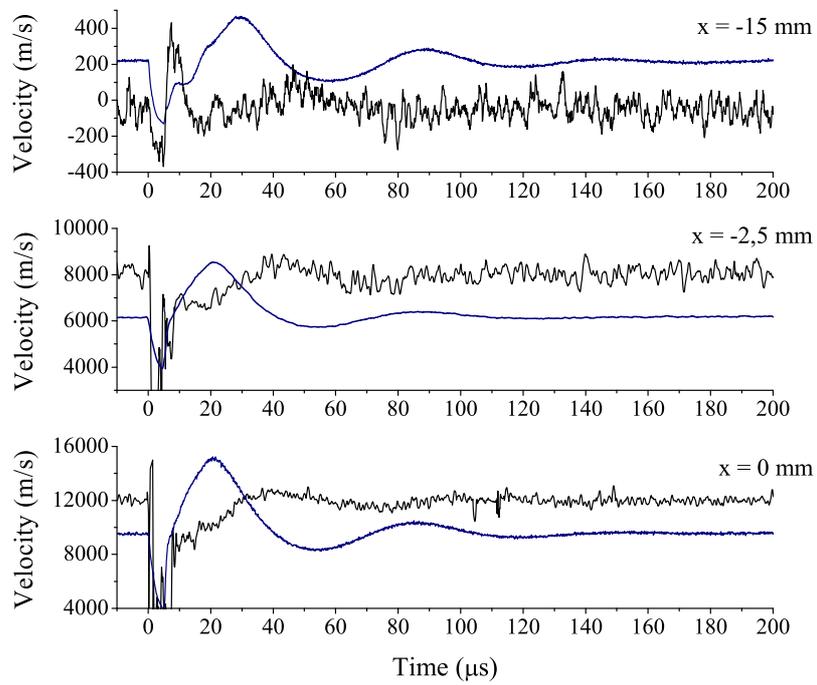}
\caption{\small{Time evolution of the Xe$^+$ ion mean axial velocity
for three on-axis positions. The time-averaged velocity measured in
a previous study at $x = -2.5$\,mm and 0\,mm is 9300\,m/s and
13100\,m/s, respectively~\cite{MAZ1}. Also shown is the profile of
the light emission at 541.9 nm (blue line).}} \label{fig:VELOCITY}
\end{figure}

\begin{figure}[t]
\centering
\includegraphics[width=12cm]{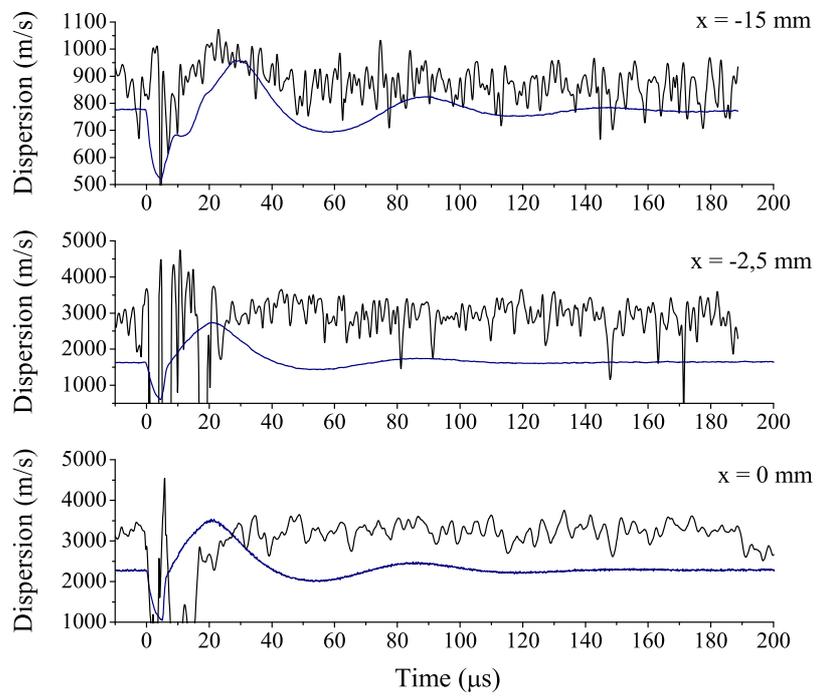}
\caption{\small{Time evolution of the Xe$^+$ ion axial velocity
dispersion ($p$ parameter) for three positions along the channel
axis of the PPS$\circledR$X000 Hall thruster. Also shown is the
profile of the light emission at 541.9 nm (blue line).}}
\label{fig:DISPERSION}
\end{figure}

The time evolution of various averaged quantities is plotted in
Fig.~\ref{fig:DENSITY},~\ref{fig:VELOCITY} and ~\ref{fig:DISPERSION}
for three positions along the channel axis, respectively $x =
-15$\,mm, $x = -2.5$\,mm and $x = 0$\,mm. The relative metastable
Xe$^+$ ion density is given by the area of the VDF. The mean
velocity and the velocity dispersion are computed from,
respectively, the first and the second order moments of the velocity
distribution. The velocity spread is in fact expressed in terms of
$p$ parameter~\cite{GAWRON,MAZ1}. The latter reads:
\begin{equation}
\label{disperson} p = 2\,\sqrt{2Ln(2)} \times \sigma \approx 2.335
\times \sigma,
\end{equation}
where $\sigma$ is the standard deviation. The quantity $p$ is equal
to the FWHM in the case of a Gaussian profile.

As can be seen in Fig.~\ref{fig:DENSITY}, the ion density is
oscillating in time with a frequency of about 16\,kHz whatever the
position. Changes in the ion density are connected with temporal
characteristics of the ionization rate. The latter is driven by a
prey-predator kind of process between atoms and charged
particles~\cite{CHOUEIRI,MODEL1,MODEL2}. Besides, Langmuir probe
measurements have shown that the electron density periodically
varies in time with the discharge current oscillation
frequency~\cite{LUC}. The fact that in Fig.~\ref{fig:DENSITY}, the
first maximum of the ion density waveform is shifted in time with
respect to the first maximum of the 541.9\,nm line intensity profile
for $x = -2.5$\,mm and $x = 0$\,mm is an experimental artefact due
to the truncation of the measured VDF.

The time evolution of the mean velocity is shown in
Fig.~\ref{fig:VELOCITY}. At $x$ = -15 mm, the velocity slightly
oscillates around zero as the observation point is outside the
acceleration layer. At $x = -2.5$\,mm and $x = 0$\,mm, the mean
velocity oscillates around a constant value, which corresponds with
a reasonably good agreement to the value obtained by means of
time-averaged LIF spectroscopy at the same location~\cite{MAZ1}. The
small difference is due to the power-switch disturbance. As shown in
Fig.~\ref{fig:VELOCITY}, the velocity changes in time are however
small. In terms of kinetic energy the largest gap is 10\,eV at
-2.5\,mm, respectively 25\,eV at the channel exit plane. Energy
variation of a few tens of eV during a low-frequency current
oscillation cycle was recorded in the plume far-field of a SPT100-ML
Hall thruster using a repulsing potential analyzer~\cite{ANDRE}.
Hybrid fluid/kinetic models also predict low-amplitude periodical
variation of the ion velocity~\cite{MODEL1,MODEL2} with a strong
correlation between discharge current and ion velocity temporal
behavior. The observed oscillations of the mean ion axial velocity
can solely be explained by a back and forth motion of the ionization
and acceleration layers and/or a change in the electric field
magnitude, as already suggested by the analysis of ion velocity
groups characteristics as a function of time.

As can be seen in Fig.~\ref{fig:DISPERSION}, the velocity dispersion
does not vary much in time other than at current shut off and
restart. At the thruster channel exhaust, the velocity dispersion
($p$ parameter) is almost constant with a value of 2950\,m/s. In
terms of energy spread it is 47\,eV. In previous
investigations~\cite{GAWRON,MAZ1}, it was clearly demonstrated that
in a Hall thruster environment the velocity dispersion originates
mostly in the spatial overlap between the ionization and the
acceleration layers. As the velocity dispersion stays unchanged with
time whereas the ion density and velocity do vary, one can conclude
the ionization front and the electric field distribution have a
correlated dynamics. For instance, assuming the form of the two
profiles stay unchanged, the two layers would have to move along at
once together in the axial direction. Actually, computer simulations
indicate that both the shape and the location of the ionization and
electric field profiles change with time~\cite{MODEL1}.

\section{Low-frequency electric field oscillations}
\label{SEC_EFIELD}

\begin{figure}[t]
\centering
\includegraphics[width=12cm]{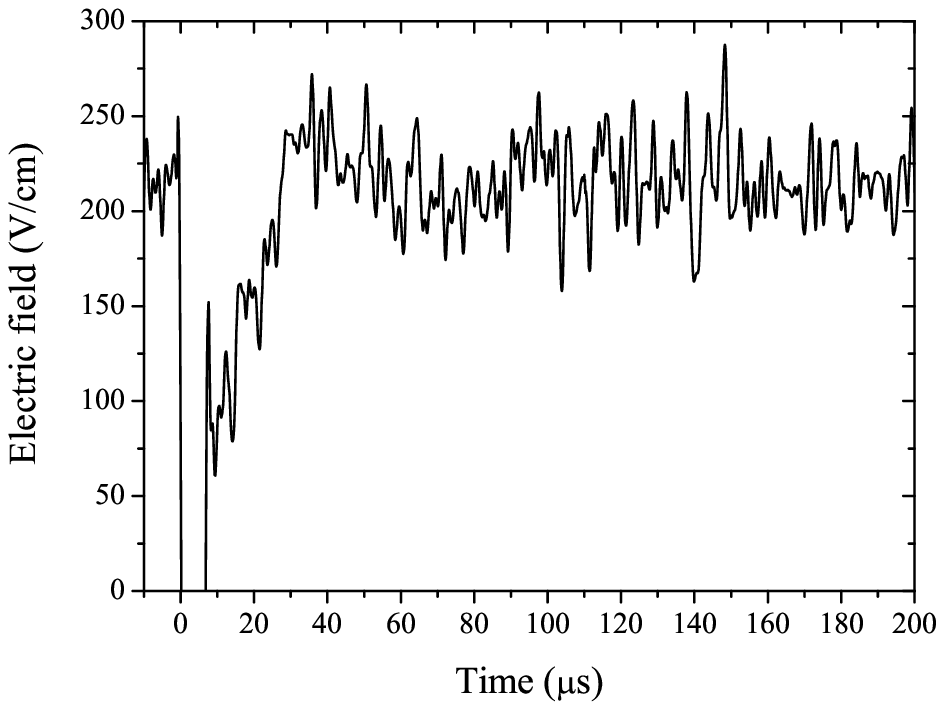}
\caption{\small{Electric field temporal characteristics ahead of the
PPS$\circledR$X000 thruster channel exhaust (between 0 and
-2.5\,mm).} The field strength is determined from the mean Xe$^+$
ion axial velocity.} \label{fig:EFIELD}
\end{figure}

The oscillation in time of the accelerating electric field can be
assessed from the time-dependent profile of the Xe$^+$ ion mean
velocity $\bar{v}$, see Fig.~\ref{fig:VELOCITY}, assuming a
collisionless medium. The electric field is computed according to
the formula:
\begin{equation}
\label{E_field} E(t) = \frac{m_{Xe^+}}{2e} \left(
\frac{\bar{v}(x_1,t)^2-\bar{v}(x_2,t)^2}{d_{x_1x_2}} \right),
\end{equation}
where $x$ is the position and $d$ refers to the distance. The
time-dependent electric field is plotted in Fig.~\ref{fig:EFIELD}
for the area that ranges between -2.5\,mm and the exit plane. The
electric field is found to oscillate with a period $T \approx
90\,\mu$s, i.e. $f \approx 11$\,kHz, around a mean value of about
215\,V/cm disregarding the power-off period. This value is close to
the value of 245\,V/cm found by way of time-averaged laser
spectroscopy~\cite{MAZ1}. The amplitude of the field oscillations
is, however, relatively weak. Over the first free oscillation that
extends from $t = 75\,\mu$s until $t = 165\,\mu$s the amplitude
varies at most from 190\,V/cm to 240\,V/cm, therefore, the electric
field variation is in the range $\pm$10\,\% ahead of the thruster
channel exhaust. Finally, in Fig.~\ref{fig:EFIELD}, one can also
notice the slow rise of the electric field magnitude from 100\,V/cm
at 10\,$\mu$s to 250\,V/cm at 30\,$\mu$s. This temporal evolution is
likely to be connected with the time it takes for the establishment
of an equilibrium state for both the discharge and the Hall
currents.

\section{Conclusions}
\label{SEC_CONCLU}

The examination of the temporal characteristics of the Xe$^+$ ion
velocity distribution function in the magnetized discharge of a Hall
effect thruster reveal the complex dynamics of the ionization and
the acceleration processes. Three results are especially of great
interest. First, the ionization profile and the electric field
distribution are unstationary. Second, the ionization and
acceleration layers vary in time in such a way that their spatial
overlap stays almost unchanged as the ion velocity spread does not
change much after ignition. Third, the amplitude of the
low-frequency electric field oscillations ahead of the channel exit
plane is weak.

Even though this study has brought new facts about the physics of a
Hall effect thruster, it appears necessary to carry on this type of
experiments aiming at building-up of a larger set of data. The
latter is actually necessary to improve our understanding of the
time and space evolution of the electric field in a Hall thruster.
Measurements of the time-dependent ion VDF must be performed with a
better signal-to-noise ratio at many locations along the thruster
channel centerline. Measurements must also be carried out for
several thruster operating conditions and various power levels.
Finally, experimental results must be critically compared with
numerical outcomes of PIC models of Hall thruster behavior.

\section*{Acknowledgements}
Works are performed in the frame of the joint-program
CNRS/CNES/SNECMA/Universities 3161 entitled ``\emph{Propulsion par
plasma dans l'espace}''. They are also financially supported by the
French National Research Agency in the frame of the 06-BLAN-0171
{\it TELIOPEH} project.



\clearpage

\clearpage

\clearpage

\clearpage

\clearpage

\begin{thebibliography}{99}
\bibitem{JPL} R. H. Frisbee, J. Propul. Power {\bf 19}, 1129
(2003).
\bibitem{MARTINEZ} M. Martinez-Sanchez, J. E. Pollard, J.
Propul. Power {\bf 14}, 688 (1998)
\bibitem{ZHURIN} V. V. Zhurin, H. R. Kaufmann and R. S. Robinson, Plasma Sources Sci. Technol. {\bf
8}, R1 (1999).
\bibitem{KIM} V. Kim, J. Propul. Power {\bf 14}, 736 (1998)
\bibitem{GASCON} N. Gascon, M. Dudeck and S. Barral, Phys. Plasmas {\bf
10}, 4123 (2003).
\bibitem{CHOUEIRI} E. Y. Choueiri, Phys. Plasmas {\bf 8}, 1411
(2001).
\bibitem{JACEK} J. Kurzyna, S. Mazouffre, A. Lazurenko, L. Albarède, G. Bonhomme, K. Makowski, M. Dudeck and Z. Peradzynski,
Phys. Plasmas {\bf 12}, 123506 (2005).
\bibitem{BOEUF} J. P. Boeuf and L. Garrigues, J. Appl. Phys {\bf 84},
3541 (1998).
\bibitem{VANESSA} V. Vial, S. Mazouffre, M. Prioul, D. Pagnon and A. Bouchoule, IEEE Trans. Plasma. Sci. {\bf 33}, 524
(2005).
\bibitem{LUC} L. Albar\'ede, S. Mazouffre, A. Bouchoule and M.
Dudeck, Phys. Plasmas {\bf 13}, 063505 (2006).
\bibitem{ANDRE} A. Bouchoule et al, Plasma Sources Sci.
Technol. {\bf 10}, 364 (2001).
\bibitem{HARGUS} W. A. Hargus Jr and M. A. Cappelli,Appl. Phys. B {\bf
72}, 961 (2001).
\bibitem{HARGUS2} W. A. Hargus Jr and C. S. Charles, J.
Propul. Power {\bf 24}, 127 (2008).
\bibitem{GAWRON} D. Gawron, S. Mazouffre, N. Sadeghi and A. H\'eron, Plasma Sources Sci.
Technol. {\bf 17}, 025001 (2008).
\bibitem{MAZ1} S. Mazouffre, D. Gawron, V. Kulaev and N. Sadeghi, IEEE Trans. Plasma
Sciences {\bf 36}, 1967 (2008).
\bibitem{NADER} B. Pellissier and N. Sadeghi, Rev. Sci. Instrum. {\bf
67}, 3405 (1996).
\bibitem{IEPC2003} V. Vial, A. Lazurenko, A. Bouchoule, M. Prioul, in \emph{Proceedings of the 28th International Electric Propulsion
Conference}, Toulouse, France, IEPC paper 2003-220.
\bibitem{MODEL1} J. Bareilles, G. J. M. Hagelaar, L. Guarrigues, C.
Boniface, J. P. Boeuf, N. Gascon, Phys. Plasmas {\bf 11}, 3035
(2004).
\bibitem{MODEL2} J.M. Fife, M. Martinez-Sanchez, J. Szabo, in \emph{Proceedings of the 33rd Joint Propulsion Conference and Exhibit}, Seattle, WA,
AIAA paper 1997-3052.
\end{thebibliography}
\end{document}